# Gradient-based adaptive interpolation in super-resolution image restoration

Jinyu Chu[1], Ju Liu[1], Jianping Qiao[1], Xiaoling Wang[1] and Yujun Li[2]
1. School of Information Science and Engineering, Shandong University, Jinan, 250100, P.R.China
2. Hisense Group, Qingdao, 266071, P.R.China
E-mail: cjy2020@mail.sdu.edu.cn   juliu@sdu.edu.cn

**Abstract**

*This paper presents a super-resolution method based on gradient-based adaptive interpolation. In this method, in addition to considering the distance between the interpolated pixel and the neighboring valid pixel, the interpolation coefficients take the local gradient of the original image into account. The smaller the local gradient of a pixel is, the more influence it should have on the interpolated pixel. And the interpolated high resolution image is finally deblurred by the application of wiener filter. Experimental results show that our proposed method not only substantially improves the subjective and objective quality of restored images, especially enhances edges, but also is robust to the registration error and has low computational complexity.*

## 1. Introduction

Super-resolution (SR) is the process of combining a sequence of noisy low resolution (LR) images in order to produce a higher resolution image or sequence. SR images can offer more details that may be critical in various applications, such as surveillance system, medical imaging, satellite image, etc. It has become an active research area in recent years.

Typically, there are two categories of algorithms in SR restoration. One is learning-based algorithm, and the other one is reconstruction-based algorithm. Reconstruction-based algorithm can be classified into iterative algorithm (e.g. maximum a posteriori/ projection onto convex sets (MAP/POCS) [1]) and non-iterative algorithm, like non-uniform interpolation [2]. This paper mainly focuses on the non-iterative algorithm since it has lower complexity and the SR restoration can be realized in real time for practical applications. Many non-iterative algorithms are expanded from single image interpolation. At present, single image interpolations have pixel replication, bilinear, spline interpolation [3], new edge-directed interpolation method (NEDI) [4] and gradient-based adaptive interpolation [5], etc., the visual results of these interpolation methods all suffer from unacceptable effects (e.g. blurring, aliasing, blocking ) to some extent, especially in edge areas of the image. However, among these interpolations, gradient-based adaptive interpolation is the most promising method to deal with the SR restoration problem.

In this paper, we propose a method based on gradient-based adaptive interpolation. Our method not only considers the distance between the interpolated pixel and the neighboring valid pixel, but also takes into account the local gradient of the original image. Firstly, we utilize the frequency domain registration algorithm to estimate the motions of the low resolution images and map the low resolution images to the uniform high resolution grid, and then the gradient-based adaptive interpolation is used to form a high resolution image. Finally, wiener filter is applied to reduce the effects of blurring and noise. Simulations show the effectiveness and robustness of our method.

This paper is organized as follows. Section 2 presents the degradation model. In Section 3, our proposed super-resolution algorithm is introduced. Experimental results are given in Section 4. Finally, we conclude with summary in Section 5.

## 2. Degradation model

Super-resolution restoration can be considered as a sparse linear optimization problem. Given $N$ low-resolution images $\mathbf{Y}_1 ... \mathbf{Y}_N$, the imaging process of $\mathbf{Y}_k$ from the super-resolution image $\mathbf{X}$ can be formulated by [6]:

$$\vec{y}_k = \mathbf{D}_k \mathbf{H}_k \mathbf{W}_k \vec{x} + \vec{n}_k \qquad (1)$$

Where $\vec{x}$ denotes a vector-wise of high resolution $\mathbf{X}$, $\vec{y}_k$ denotes a vector-wise of the $k$-th low resolution image $Y_k$. $\vec{n}_k$ is the corresponding normally distributed additive noise $N_k$ in the $k$-th low-resolution image. $D_k$, $H_k$ and $W_k$ denote the decimation matrix, the blurring matrix, and the geometric warp matrix, respectively. By considering all the frames and by stacking the vector equations, we obtain a matrix-vector formula:

$$\begin{bmatrix} \vec{y}_1 \\ \vdots \\ \vec{y}_N \end{bmatrix} = \begin{bmatrix} D_1 H_1 W_1 \\ \vdots \\ D_N H_N W_N \end{bmatrix} \vec{x} + \begin{bmatrix} \vec{n}_1 \\ \vdots \\ \vec{n}_N \end{bmatrix} \quad (2)$$

## 3. Super-resolution restoration algorithm based on gradient adaptive interpolation

The basic idea of the gradient-based adaptive interpolation is that the interpolated pixel value is influenced by the local gradient of a pixel, especially in the edge areas of the image. Actually the smaller the local gradient of a pixel is, the more influence it should have on the interpolated pixel.

Inspired by this idea, we propose a novel SR restoration algorithm based on gradient-based adaptive interpolation. The proposed method involves three sub-tasks: registration, fusion and deblurring. Firstly we utilize the frequency domain registration algorithm [7] to estimate the motions of the low resolution images. According to the motions the low resolution images are mapped to the uniform high resolution grid, and then the gradient based adaptive interpolation is used to form a high resolution image. Finally, wiener filter is applied to reduce the effects of blurring and noise caused by the system.

The details of the proposed algorithm are described as follows:

Step 1: Take the reference LR image (e.g. frame 1 in Fig.1) and place it on the uniform high resolution grid at position [0, 0].

Step 2: Estimate the motion vectors between the reference image and the other LR images using frequency domain registration algorithm [7].

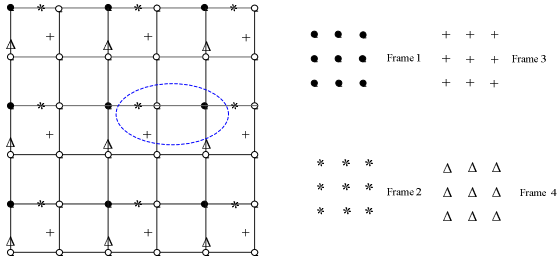

Fig.1. high resolution grid points

Step 3: Determine other low resolution images' positions relative to the uniform high resolution grid points.

Step 4: Calculate the Euclidean distances between the interpolated pixel and its neighboring pixels (with size $5\times 5$) in the high resolution grid points, rank the distances from the closest to the farthest and find the 3 nearest pixels around the interpolated one. In Fig.1, the interpolated pixel (white circle ○) and its 3 nearest pixels are illustrated in the ellipse.

Step 5: According to the gradient-based adaptive interpolation algorithm [5], compute the interpolated gray values. The interpolated point $(i, j)$ and its 3 nearest points $(i_k, j_k)$ $k=1,2,3$, the gray value $f(i,j)$ of the interpolated point $(i,j)$ is illustrated as follows:

$$f(i,j) = \frac{\sum_{k=1}^{3} W(i_k, j_k) S(i_k, j_k) f(i_k, j_k)}{\sum_{k=1}^{3} W(i_k, j_k) S(i_k, j_k)} \quad (3)$$

Where, $S(i_k, j_k) = (1 - \Delta x(i_k, j_k))(1 - \Delta y(i_k, j_k))$ is the distance function which is calculated using the vertical $\Delta x(i_k, j_k)$ and horizontal distance $\Delta y(i_k, j_k)$:

$$\begin{aligned} \Delta x(i_k, j_k) &= |i_k - i| \\ \Delta y(i_k, j_k) &= |j_k - j| \end{aligned} \quad (4)$$

and $W(i_k, j_k) = (-\mu G(i_k, j_k) + 1)^m$ is the gradient weight function, where $m$ is a positive value, $\mu$ is a positive value close to and smaller than 1, and $G(i_k, j_k)$ is the local gradient of point $(i_k, j_k)$, $G(i_k, j_k)$ is represented by the following formulation [8]:

$$G(i_k, j_k) = \frac{|f'_x(i_k, j_k)| + |f'_y(i_k, j_k)|}{2\sqrt{(f'_x(i_k, j_k))^2 + (f'_y(i_k, j_k))^2}} \quad (5)$$

For each of the pixel $(i_k, j_k)$, its vertical derivative $f'_x(i_k, j_k)$ and the horizontal derivative $f'_y(i_k, j_k)$ are evaluated by Sobel masks of size 3 (one vertical Sobel1=[-1 -2 -1;0 0 0;1 2 1] and one horizontal Sobel2=[-1 0 1; -2 0 2;-1 0 1]) in the low resolution image which $(i_k, j_k)$ belongs to.

Step 6: Wiener filter is applied to reduce the effects of blurring and noise caused by the system [9].

The gradient-based adaptive interpolation takes into account the local gradient. The smaller the local gradient of a pixel, the more influence it should have on the interpolated pixel. Assume that an interpolated pixel $\alpha$ lies exactly between two pixels $\alpha_1$ (a flat region pixel) and $\alpha_2$ (an edge pixel), which is demonstrated in Fig.2. Although $\alpha$ is closer to the pixel $\alpha_2$ in the distance, it is clear that it should be closer to $\alpha_1$ in the gray value, since $\alpha_2$ has a higher local gradient and is evidently on an edge. Therefore, in addition to considering the distance, the local gradient should also be taken into account.

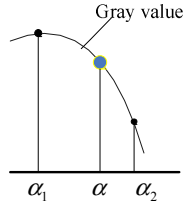

Fig.2. local gradient [5]

## 4. Experimental results

### 4.1. Standard image reconstruction

In this experiment some standard images are used to demonstrate the effectiveness of the proposed method. Four blurred, sub-sampled and noisy low-resolution images are generated through the high resolution standard images of size 256×256 and gray level 256. The Blurred Signal-to-Noise Ratio (BSNR) is 30dB, 25dB, 35dB and 30dB, respectively, decimation ratio is 2:1, we use shifts as geometric warp and the motion vectors are set as [0 0], [0 -0.8], [-0.8 -0.8] and [-0.8 0] respectively.

Table 1. PSNR comparison (dB)

| Standard image | Bilinear | NEDI | MAP\POCS | Nonuniform interpolation | The proposed method |
|---|---|---|---|---|---|
| Monarch | 20.4461 | 23.7609 | 24.5383 | 26.5278 | 27.3344 |
| Panda | 21.5227 | 25.4398 | 25.5231 | 27.9256 | 28.7383 |
| Crowd | 21.1302 | 25.0131 | 25.8514 | 27.8979 | 28.8979 |
| Lena | 22.1357 | 26.2154 | 26.1686 | 27.2086 | 27.6223 |
| Cameraman | 20.9832 | 24.1896 | 24.5246 | 25.5407 | 25.8120 |

Table 2. MSSIM comparison

| Standard image | Bilinear | NEDI | MAP\POCS | Nonuniform interpolation | The proposed method |
|---|---|---|---|---|---|
| Monarch | 0.7925 | 0.8787 | 0.9015 | 0.9280 | 0.9370 |
| Panda | 0.7319 | 0.8378 | 0.8475 | 0.8981 | 0.9059 |
| Crowd | 0.6699 | 0.8059 | 0.8444 | 0.9061 | 0.9261 |
| Lena | 0.6783 | 0.7874 | 0.8023 | 0.8482 | 0.8544 |
| Cameraman | 0.7173 | 0.8029 | 0.8228 | 0.8486 | 0.8577 |

Table 1 and Table 2 show PSNR and MSSIM comparisons of different methods. It can be seen that the proposed method can offer noticeable better results, especially for images with more edge information, such as Monarch, Panda and Crowd.

### 4.2. Real image reconstruction

Four low-resolution real images are shown in Fig.3. Take the first image as the reference image and the other are aligned into the reference image using the frequency domain registration algorithm. The shifts estimation parameters are [0 0], [-0.3229 -0.5369], [-0.4476 -0.7661] and [-0.5446 -0.0983], respectively. Comparison results are shown in Fig.4 and the local area zoom-in comparisons are shown in Fig.5.

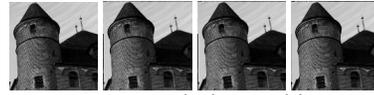

Fig.3. Low-resolution real images

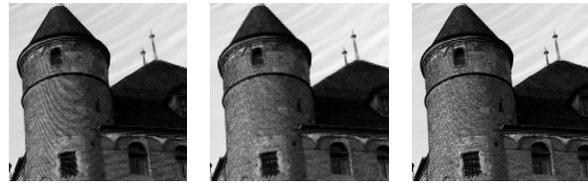

(a) NEDI (b) Non-uniform (c) Our method
Fig.4. Restored images comparison

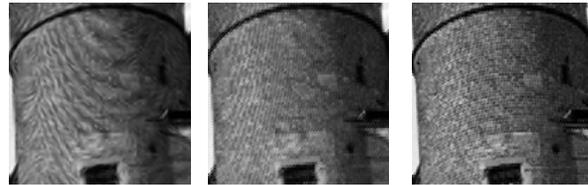

(a) NEDI (b) Non-uniform (c) Our method
Fig.5. Zoom-in restored images comparison

It can be seen that the proposed method preserves the edges information better and generates reconstructed images with higher visual quality.

### 4.3. Robustness to registration error

The parameter settings of the simulation are the same as Section 4.1. We use shifts as geometric warp, real motion vectors are shown as case 1 in Table.3. In order to show the robustness to registration error, two kinds of motion vectors are set for the comparisons as case 2 (smaller registration error) and case3 (larger registration error) in Table.3. Simulation results are shown in Fig.6 and Fig.7.

Table 3. The inaccurate sub-pixel motion comparison

| Motion | $[n_1^1, n_2^1]$ | $[n_1^2, n_2^2]$ | $[n_1^3, n_2^3]$ | $[n_1^4, n_2^4]$ |
|---|---|---|---|---|
| Case 1 | [0.0 0.0] | [0.0 -0.4] | [-0.4 -0.4] | [-0.4 0] |
| Case 2 | [0.0 0.0] | [0.0 -0.3] | [-0.2 -0.3] | [-0.2 0] |
| Case 3 | [0.0 0.0] | [0.0 -0.1] | [-0.1 -0.1] | [-0.1 0] |

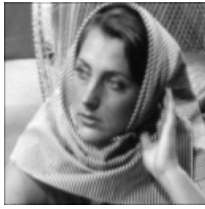 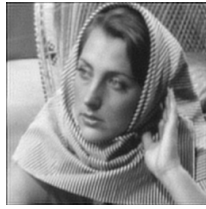

(a) MAP/POCS (PSNR=24.5649dB)      (b) Our method (PSNR=25.8678dB)

Fig.6. Restored images comparison of case 2

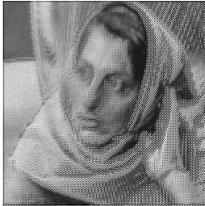 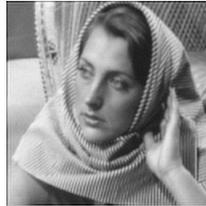

(a) MAP/POCS (PSNR=10.4135dB)      (b) Our method (PSNR=25.8145dB)

Fig.7. Restored images comparison of case 3

It can be seen that when registration error increases, the performance of MAP/POCS method will decrease. However, our proposed method can still obtain better results from either subjective feeling or objective appreciation based on PSNR. Hence, the proposed method is robust to registration error.

Moreover, the operation time of MAP/POCS method is 16.03s, while the proposed method is 6.43s. Thus the proposed method is faster than iterative method and is more attractive for real time applications.

## 5. Conclusions

In this paper, we propose a novel super-resolution method based on gradient-based adaptive interpolation. Firstly, the relative motions of the LR images are estimated, and then gradient adaptive interpolation is used to form an interpolated high resolution image, the interpolation is adjusted by the local gradient properties. Finally, wiener filter is applied to remove blurring and noise. Simulations indicate that the proposed method can obtain well-visual restored SR images, and have robustness to registration error. What's more, the proposed method has low computational complexity and hence is suitable for real time applications.

## Acknowledgment

This work is supported by Program for New Century Excellent Talents in University Education Ministry of China (NCET-05-0582), the Excellent Youth Scientist Award Foundation of Shandong Province (No. 2007BS01023; No.2007BS01006), the Specialized Research Fund for the Doctoral Program of Higher Education (No.20050422017) and Natural Science Foundation of Shandong Province (No.Y2007G04). The corresponding author is Ju Liu (juliu@sdu.edu.cn).